\documentclass[11pt,twoside]{article}
\usepackage{CAGN2016}
\usepackage{graphicx}

\usepackage[T1]{fontenc} 

\usepackage{latexsym}
\usepackage{verbatim}

\newcommand{\hi}{H\thinspace\textsc{i}}
\newcommand{\foiii}{[O\thinspace\textsc{iii}]}

\newcommand{\cii}{C\thinspace\textsc{ii}}
\newcommand{\oi}{O\thinspace\textsc{i}}
\newcommand{\oii}{O\thinspace\textsc{ii}}
\newcommand{\neii}{Ne\thinspace\textsc{ii}}
\newcommand{\hii}{H\thinspace\textsc{ii}}

\newcommand{\hei}{He\thinspace\textsc{i}}

\begin{document}

\vskip 1.0cm
\markboth{C.~Esteban et al.}{Abundance discrepancy in {\hii} regions}
\pagestyle{myheadings}
%
%
\vspace*{0.5cm}
\parindent 0pt{Invited Review}


\vspace*{0.5cm}
\title{The abundance discrepancy in {\hii} regions}

\author{C.~Esteban$^{1, 2}$, L.~Toribio San Cipriano$^{1, 2}$ and J. Garc\'ia-Rojas$^{1, 2}$}
\affil{$^1$Instituto de Astrof\'isica de Canarias, E-38200 La Laguna, Tenerife, Spain\\
$^2$Departamento de Astrof\'isica, Universidad de La Laguna, E-38206, La Laguna, Tenerife, Spain}

\begin{abstract}
In this paper we discuss some results concerning the abundance discrepancy problem in the context of {\hii} regions. We discuss the behavior of the abundance discrepancy factor (ADF) for different objects and ions. There are evidences that stellar abundances seem to agree better with the nebular ones derived from recombination lines in high-metallicity environments and from collisionally excited lines in the low-metallicity regime. Recent data point out that the ADF seems to be correlated with the metallicity and the electron temperature of the objects. These results open new ways for investigating the origin of the abundance discrepancy problem in {\hii} regions and in ionized nebulae in general. 
\end{abstract}

\section{Introduction}
\label{intro}

The emission-line spectrum of {\hii} regions can be detected and analyzed even at very large distances. They are essential probes to measure distances of extragalactic objects, the 
intensity and properties of star formation processes and the chemical composition of the interstellar medium. In fact, {\hii} regions provide most of our current knowledge about the 
chemical composition of the Universe. The spectra of nebulae contain emission lines of several ionization species of chemical elements that are present in the ionized gas. The 
analysis of emission-lines permits to determine the abundance -- the ratio of the number of atoms of a given element with respect to those of H -- of important elements as He, 
C, N, O, Cl, $\alpha$-elements as Ne, S and Ar and iron-peak elements as Fe and Ni. In the next decade, the use of 30-50m aperture telescopes will probably permit to derive 
abundances of some neutron-capture s-elements as Se or Kr in the brightest {\hii} regions.   

The O abundance determined from nebular spectra is the most widely used proxy of metallicity for galaxies at different redshifts. A proper determination of the physical conditions -- 
electron temperature, $T_e$, and density, $n_e$ -- is necessary for obtaining reliable and accurate metallicities.  Many of the emission lines in nebular spectra are excited by collisions with 
free electrons -- the so-called collisionally excited lines (CELs) -- and are forbidden by the selection rules. CELs can be very bright due to the high electron temperatures and low densities of ionized nebulae. In addition, there are also bright recombination lines of {\hi} -- Balmer series in the optical -- and {\hei}. In the last two decades, our group has developed a long-term research project devoted to measure very faint pure recombination 
lines (RLs) of heavy-element ions in {\hii} regions from high- and intermediate-resolution spectroscopical data. The optical-NIR spectral range  contains RLs of {\oi}, {\oii}, {\cii} and 
{\neii}, which intensities are of the order of 0.0001 to 0.001 times that of H$\beta$. Due to their faintness, these lines can only be detected and measured in bright nebulae using 
large aperture telescopes.

For several decades it has turned out that abundances of heavy-element ions determined from the standard method based on the intensity ratios of co\-lli\-sio\-na\-lly excited lines (CELs) are systematically lower than those derived from the faint RLs emitted by the same ions. This fact, usually known as the {\em abundance discrepancy} (AD) problem is quantified by the abundance discrepancy factor (hereafter ADF), defined as the difference between the logarithmic abundances derived from RLs and CELs:
\begin{equation}
 {\rm ADF}({\rm X}^{\rm i}) = {\rm log}({\rm X}^{\rm i}/{\rm H}^+)_{\rm RLs} - {\rm log}({\rm X}^{\rm i}/{\rm H}^+)_{\rm CELs}.
 \label{ADF_equation}
\end{equation} 
In Galactic and extragalactic {\hii} regions, the ADF of O$^{2+}$ -- ADF(O$^{2+}$) -- is between 0.10 and 0.35 dex (e.g. Garc\'ia-Rojas \& Esteban 2007; L\'opez-S\'anchez et al. 2007; Esteban et al. 2009, 2014; Toribio San Cipriano et al. 2016, 2017) and can be even much larger in some planetary nebulae (PNe, see paper by Garc\'ia-Rojas in this proceedings). It is important to remark that other ions with optical RLs: C$^{2+}$, Ne$^{2+}$, and O$^+$ give values of their ADFs similar to those obtained for the ADF(O$^{2+}$) for the same object (Garc\'{\i}a-Rojas \& Esteban 2007; Toribio San Cipriano et al. 2017), although the data for those ions are more limited and difficult to obtain (see Table~\ref{table}). It is clear that changes or uncertainties in the face value of metallicity we adopt for celestial bodies may have a major impact in many fields of Astrophysics as the ingredients of chemical evolution models and predicted stellar yields (e.g. Carigi et al. 2005),  the luminosity --and mass-- metallicity relations for local and high-redshift star-forming galaxies (e.g. Tremonti et al. 2004), the calibration of strong-line methods for deriving the abundance scale of extragalactic {\hii} regions and star-forming galaxies at different redshifts (e.g. Pe\~na-Guerrero et al. 2012, L\'opez-S\'anchez et al. 2012) or the determination of the primordial helium (e.g. Peimbert 2008), among others.

\begin{table*}
\caption{ADF for different ions and several H\thinspace{\sc ii} regions (in dex).}\label{ADFs}
\label{table}
\begin{tabular}{lccccc}
\hline
& \multicolumn{2}{c}{Milky Way} & LMC & SMC & \\
Ion & Orion Neb. & M8 & 30 Dor & N66C & NGC~5253 \\
\hline
O$^+$ & +0.39$\pm$0.20 & +0.14$\pm$0.09 &  +0.26$\pm$0.13 & ... & ... \\
O$^{2+}$ & +0.14$\pm$0.01 & +0.37$\pm$0.04 &  +0.21$\pm$0.02& +0.35$\pm$0.13 & +0.25$\pm$0.16 \\
C$^{2+}$ & +0.40$\pm$0.15 & +0.54$\pm$0.21& +0.25$\pm$0.21& +0.45$\pm$0.15 & +0.41$\pm$0.25 \\
Ne$^{2+}$ & +0.26$\pm$0.10 & ...& ...& ...& ...\\
Ref.& 1& 2& 3& 4& 5\\
\hline
\end{tabular}
      \\
      1- Esteban et al. (2004); 2- Garc\'{\i}a-Rojas et al. (2007); 3- Peimbert (2003); \\
      4- Toribio San Cipriano et al. (2017); 5- L\'opez-S\'anchez et al. (2007). 
\end{table*}

The origin of the AD is still unknown, but several hypotheses have been proposed. In this paper we will limit our discussion to the case of {\hii} regions. For 
PNe the situation is more complex because they can contain material with different physical conditions and composition (see Peimbert \& Peimbert 2006; Garc\'ia-Rojas, this 
proceedings). The first hypothesis for explaining the AD was formulated by Torres-Peimbert et~al. (1980). They propose that the AD is produced by spatial fluctuations of 
electron temperature, $t^2$, in the ionized gas, in the form originally proposed by Peimbert (1967). However, these temperature variations cannot be reproduced by 
chemically homogeneous photoionization models and 
different mechanisms have been invoked to explain their presence in ionized nebulae (see Esteban 2002; Peimbert \& Peimbert 2006). According to this scenario, RLs should 
provide the true abundances because their emissivities are much less dependent on temperature that those of CELs. In fact, in the typical range of $T_e$ of {\hii} regions, the 
emissivities of the RLs of heavy-element ions are almost identical to that of H$\beta$ and therefore abundances derived from RLs are practically independent on $T_e$. Nicholls 
et~al. (2012) have proposed that a $\kappa$-distribution of electron energies -- departure from the Maxwell-Boltzmann distribution -- can be operating in nebulae and produce the 
AD, being also the determinations based on RLs the most reliable ones. However, some recent works have seriously questioned this hypothesis (e. g. Zhang et~al. 2016; Ferland et~al. 
2016). A third hypothesis was proposed by Stasi\'nska et~al. (2007) and is based on the existence of semi-ionized clumps embedded in the ambient gas of {\hii} regions. These 
hypothetical clumps -- unmixed supernova ejecta -- would be strong RL emitters, denser, cooler and more metallic than the ambient nebular gas. Assuming  this scenario, abundances 
derived from RLs and CELs should be upper and lower limits, respectively, to the true ones, though those from CELs should be more reliable. There is even a last scenario outlined 
by Tsamis et al. (2011), which involves the presence of high-density clumps but without abundance contrast between the components. 

Until now, the only work where the AD problem in a sample of {\hii} regions has been discussed in length is that by Garc\'ia-Rojas \& Esteban (2007). They found that the ADF is 
fairly constant and of order 2 in a limited sample of Galactic and extragalactic {\hii} regions. In addition, they did not find correlations between the ADF(O$^{2+}$) and the O/H, O$^{2+}$/H$^+$ 
ratios, the ionization degree, $T_e$, FWHM of several bright emission-lines, and the effective temperature of the main ionizing stars within the observational uncertainties. On the 
contrary, they found that  ADF seems to be slightly dependent on the excitation energy of the levels that produce the RLs, a fact that is consistent with the predictions of the 
classical temperature fluctuations or $\kappa$-distributions hypotheses. 

\section{Comparisons between nebular and stellar abundances}
\label{comparisons}

\begin{figure}  
\begin{center}
\includegraphics[angle=0, height=7.0cm]{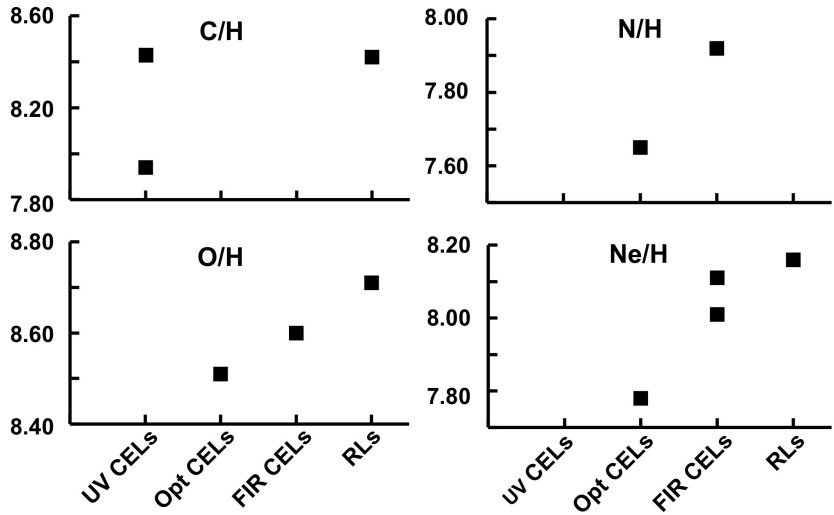}
\caption{Abundances of C, N, O, and Ne of the Orion Nebula obtained using different kinds of emission-lines. We represent ultraviolet, optical and far infrared (UV, Opt and FIR, 
respectively) collisionally excited lines (CELs) and optical recombination lines (RLs). See text for references.}
\label{abund_orion_1}
\end{center}
\end{figure}

If CELs and RLs give different abundances, in which lines shall we trust? There are some observational evidences that can give us  clues about this important question. In 
Figure~\ref{abund_orion_1} we represent the abundances of 4 representative elements -- C, N, O, and Ne -- for which abundances have been derived from different kinds of lines in 
the Orion Nebula: CELs in the ultraviolet (UV, Walter et~al. 1992;  Tsamis et~al. 2011), optical (Esteban et~al. 2004) and far infrared (FIR, Simpson et~al. 1986; Rubin et~al. 2011), 
and optical RLs (Esteban et~al. 2004). Except in the case of C -- for which there are two discrepant independent abundance determinations from UV CELs -- there is a tendency to 
increase the abundance when we move to the right in the diagrams, in the sense that the lines which intensity is less dependent on $T_e$ give higher abundances. This 
tendency agrees qualitatively with the predictions if the AD is produced by the temperature fluctuations or $\kappa$-distribution hypotheses.

\begin{figure}  
\begin{center}
\includegraphics[angle=0, height=7.0cm]{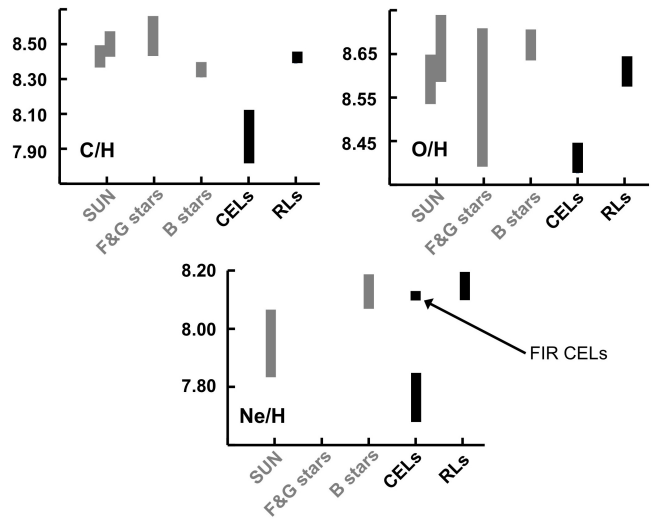}
\caption{Comparison of abundances of C, O, and Ne of the Orion Nebula determined from CELs and RLs and those of the Sun and young F\&G and B-type stars of the Solar Neighbourhood (the data for non-nebular objects are represented with grey symbols). The size of the bars is proportional to their uncertainty. See text for references.}
\label{abund_orion_2}
\end{center}
\end{figure}

In Figure~\ref{abund_orion_2} we compare the abundances of C, O and Ne of the Orion Nebula -- using UV or optical CELs and RLs -- with those of other objects of the Solar 
Neighbourhood for which absolute abundances can be derived. We include the Sun (Asplund et~al. 2009; Caffau et~al. 2008, 2009), young F\&G stars (Sofia \& Meyer 2001), and 
B-type stars (Nieva \& Sim\'on-D\'iaz 2011). The size of the bars is proportional to their uncertainty. For C/H we can see that the solar and stellar values are quite consistent with the 
values obtained for the Orion Nebula using RLs, a similar trend is also found for O/H. However, we would expect some depletion into dust for both elements in the Orion Nebula, of 
the order of about 0.1 dex according to Esteban et al. (2004). Ne is a noble gas and no dust depletion is expected for this element and the results also indicate that the stellar and 
solar Ne abundances are more consistent with the Ne/H ratios determined from RLs in the Orion Nebula. It is interesting to note that the accurate determination of Ne/H obtained by 
Rubin et al. (2011) from FIR CELs observed with {\em Spitzer} is consistent with the RLs values. This result is also consistent with the predictions of the temperature fluctuations 
or $\kappa$-distribution hypotheses considering the low temperature dependence of the emissivity of FIR CELs. From the results presented in figures~\ref{abund_orion_1} 
and \ref{abund_orion_2}, one can conclude that nebular abundances determined from RLs are more consistent with those of the Sun and other stellar objects 
of the  Solar Neighbourhood than those determined from CELs. 

Since 2002, our group has measured {\cii} and {\oii} RLs in several tens of extragalactic {\hii} regions (Esteban et~al. 2002, 2009, 2014; L\'opez-S\'anchez et al. 2007; Toribio San 
Cipriano 2016, 2017). In several of those papers we have compared our nebular abundances derived from CELs and RLs with those determined in early B-type stars located in 
their vicinity. The O abundance of early B-type stars -- which is not expected to be affected by stellar evolution effects -- should reflect the present-day chemical composition of the 
interstellar material in the regions where they are located. The results for extragalactic objects are not always consistent with the aforementioned ones discussed for the Orion Nebula. 
Toribio San Cipriano et al. 
(2016) compared O abundances derived from CELs and RLs in {\hii} regions of the spiral galaxies M33 and NGC300 with the O abundances derived in B supergiants by Urbaneja 
et al. (2005a, 2005b). These authors found that the stellar O/H ratios are in better agreement with the nebular abundances calculated using RLs in the case of M33, while in the 
case of NGC 300 the agreement is better with CELs. More recently, Toribio San Cipriano et al. (2017) make a similar comparison in the case of the Magellanic Clouds. They find 
that the O abundances determined by Hunter et al. (2009) for B-type stars in young clusters of the LMC and SMC are more consistent with the values obtained from CELs for the 
associated {\hii} regions. 

\begin{figure}  
\begin{center}
\includegraphics[angle=0, height=5.0cm]{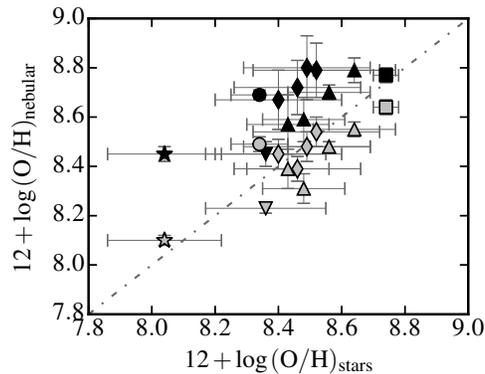}
\caption{Comparison of the O abundance determined from young supergiant stars with those obtained from {\hii}  regions located in the same star-forming region or at the same 
galactocentric distance in the same galaxy. Squares: Orion Nebula and its associated cluster; circles: N11 in the LMC; stars: N66 in the SMC;  down-facing triangles: NGC6822; 
triangles: M33; diamonds: NGC 300. Grey and black symbols represent nebular CELs-based and RLs-based O/H ratios, respectively. The dot-dashed line represents the 1:1 relation. 
Figure taken from Toribio San Cipriano et~al. (2017).}
\label{neb_stars}
\end{center}
\end{figure}

Bresolin et al. (2016) suggested that nebular abundances determined either from CELs or RLs can display different levels of agreement with the B supergiant determinations in 
different galaxies, depending on the metallicity regime. They proposed that the RL-based nebular metallicities agree with the stellar metallicities better than the CELs in the 
high-abundance regime, but that this situation reverses at low abundance values. In Figure~\ref{neb_stars} we have made a similar exercise than in figure 11 of Bresolin et al. (2016) but 
including data for individual {\hii} regions with high quality nebular CELs and RLs determinations and compare with O/H ratios determined from young supergiant stars located in 
the same star-forming regions or galactocentric distance in the same galaxy. The squares correspond to data for the Orion Nebula and B-type stars of its associated cluster 
(Esteban et~al. 2004; Sim\'on-D\'iaz \& Stasi\'nska 2011). For the MCs we include data for the ionized gas (Toribio San Cipriano et~al. 2017) and B-type stars (Hunter et~al. 2009) 
of the star- forming regions N11 (LMC, circles) and N66 (SMC, stars). We have considered data points of several {\hii} regions of M33 (triangles) and NGC300 (diamonds) taken 
from Toribio San Cipriano et~al. (2016) and Esteban et al. (2009), the stellar abundances have been estimated from the radial O abundance gradients determined by Urbaneja 
et~al. (2005a, 2005b) from the spectra of B-type supergiants, evaluated at the galactocentric distances of each {\hii} region. The down-facing triangles represent values for the 
dwarf irre\-gu\-lar galaxy NGC 6822, the nebular data are taken from Esteban et al. (2014) and the stellar ones from the spectral analysis performed by Venn et al. (2001). Nebular 
data have been increased 0.1 dex to correct for dust depletion in all the objects. From the figure, one can conclude that while in the case of M33 and NGC 6822 both kinds of lines 
give nebular abundances consistent with the stellar ones, in the other cases we obtain contradictory results. The nebular O abundances determined from RLs in the Orion Nebula 
are the ones consistent with stellar determinations while in the MCs and NGC 300 the determinations based on CELs are the only ones consistent with the stellar abundances. This 
seems to be qualitatively consistent with the result obtained by Bresolin et al. (2016). We need further high-quality data -- specially in the higher- and lower-metallicity regimes -- to 
confirm this tendency.

\section{ADF {\em versus} some nebular parameters}
\label{ADFvs}

\begin{figure}  
\begin{center}
\includegraphics[angle=0, height=15.0cm]{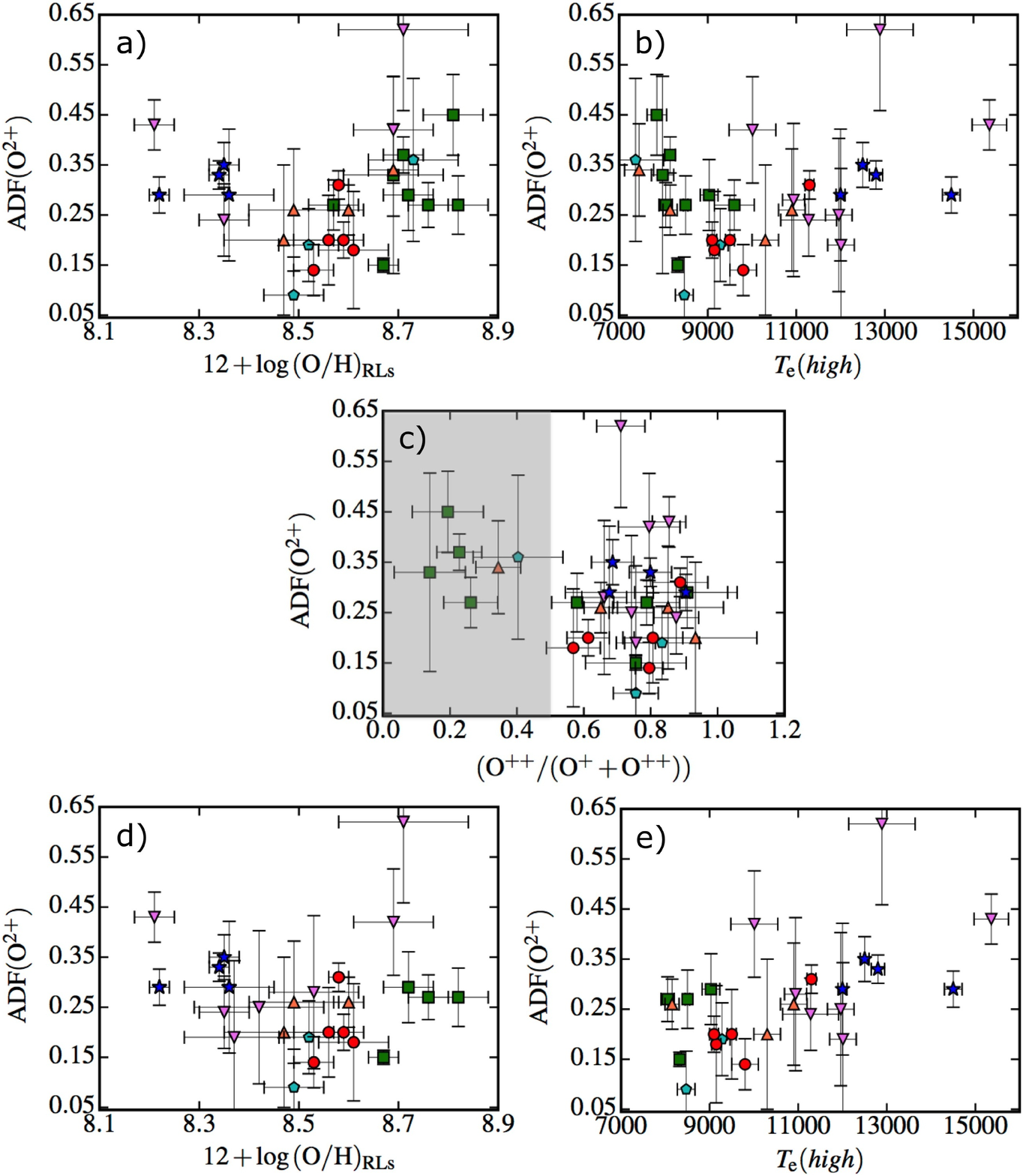}
\caption{Panels a) and b): ADF(O$^{2+}$) {\em versus} O/H ratio determined from RLs and {\em versus} $T_e$ for high ionization species for different {\hii} regions in 
different host galaxies.  Panel c): ADF(O$^{2+}$) {\em versus} O$^{2+}$/O ratio -- ionization degree -- for the same {\hii} regions. The grey area contains low-ionization objects, with 
O$^{2+}$/O $\leq$ 0.5. Panels d) and e): The same as panels a) and b) but removing the low-ionization {\hii} regions lying in the grey area of panel c). See text for 
description of the symbols.}
\label{adf_mosaic}
\end{center}
\end{figure}

Garc\'ia-Rojas \& Esteban (2007) explored the  dependence of the ADF with different properties of {\hii} regions. Those authors reported that the ADF seems to be 
independent of metallicity however, their sample was rather limited in both, the number of objects and metallicity. In Figure~\ref{adf_mosaic}, we present the behavior of the 
ADF(O$^{2+}$) with respect to O/H ratio, $T_e$, and ionization degree (O$^{2+}$/O ratio). Green squares correspond to Galactic {\hii} regions 
(Garc\'ia-Rojas \& Esteban 2007; Esteban et al. 2004, 2013); red circles and blue stars represent objects in the LMC and SMC, respectively (Toribio San Cipriano et al. 2017);  
orange triangles correspond to {\hii} regions in M33 (Toribio San Cipriano et al. 2016); cyan pentagons to objects in M101 (Esteban et al. 2009); and pink down-facing triangles to 
star-forming dwarf galaxies (L\'opez-S\'anchez et al. 2007; Esteban et al. 2014). Let's focus our attention in panel a) of Figure~\ref{adf_mosaic}, although the ADF(O$^{2+}$) values of  
many objects have large uncertainties, the distribution of the points indicates a complex behavior in the ADF(O$^{2+}$) {\em versus} O/H diagram. The ADF distribution shows a  
minimum at 12 + log(O/H) $\sim$ 8.5 and seems to increase when the O/H becomes higher or lower. This apparent ``seagull'' shape, if real, is difficult to explain. In the case of 
Galactic {\hii} regions we obtain a high dispersion of ADFs and a tendency to larger values when O/H becomes higher. Considering the closeness of the Galactic nebulae, their 
dispersion may be due to aperture effects because we tend to observe bright small areas in these objects and their ADF could not be representative of the whole nebula. As it has 
been proven in several works (e.g. Mesa-Delgado et~al. 2012, and references therein) the presence of localized high-velocity flows and/or high density clumps can produce large 
ADFs. The trend of the low-metallicity wing of the ``seagull'' shape of Figure~\ref{adf_mosaic}, seems to be clearer than the high-metallicity wing. The low-metallicity objects correspond to 
more distant objects, their spectra are obtained from apertures that encompass a large fraction of the nebula and, therefore, their observed areas are more representative of the global emission. 
Toribio San Cipriano et al. (2017) also find that the ADF(C$^{2+}$) seems to increase toward lower metallicities, reinforcing the validity of the apparent correlation between ADF(O$^{2+}$) 
and O/H shown in panel a). 

Panel b) of Figure~\ref{adf_mosaic} shows the ADF(O$^{2+}$) {\em versus} $T_e$ for the high-ionization species, basically $T_e$({\foiii}). We can see a similar seagull shape in this relation but inverted 
with respect to that shown in panel a) --. This result is not unexpected, as $T_e$ and metallicity are related because O is one of the most important coolants in ionized nebulae. The trend of a decrease of 
ADF with increasing of $T_e$({\foiii}) in {\hii} regions of near-solar metallicity and $T_e$({\foiii}) $<$ 10$^4$ K was previously reported by Rodr\'iguez \& Manso Sainz (2014). 

Panel c) of Figure~\ref{adf_mosaic} represents the ADF(O$^{2+}$) {\em versus} O$^{2+}$/O ratio -- ionization degree -- of the objects. The panel shows two main groups of objects. 
Most of them lie around O$^{2+}$/O $\sim$ 0.8 and a small band is located at O$^{2+}$/O $<$ 0.5, these last objects -- most of them Galactic {\hii} regions -- show the lowest ionization 
degrees as well as ADFs higher than the average. It is interesting to note that, in panels a) and b), this group of low-ionization objects lie precisely in the zones of high O/H ratio and 
low $T_e$ where we find the apparent trend of increasing ADFs. Panels d) and e) of Figure~\ref{adf_mosaic} show the same relations than panels a) and b) but removing the 
low-ionization objects (those with O$^{2+}$/O $<$ 0.5). In these new panels we can see that the trends of the ADF towards high metallicity and low $T_e$ are practically washed 
out. The interpretation of this behavior is not easy. In low-ionization nebulae with O$^{2+}$/O $<$ 0.5, the O/H ratio is dominated by O$^+$, and therefore the physical conditions of the 
O$^{2+}$ zone -- specially $T_e$ -- may be not representative of the whole  {\hii} region. We will further investigate this interesting result. 

Another interesting aspect of Figure~\ref{adf_mosaic} is the large ADF(O$^{2+}$) values reported in some giant {\hii} regions belonging to star-forming dwarf galaxies (pink down-facing triangles). Esteban et al. (2016) speculate that large ADFs in {\hii} regions might be produced by the presence of shocks due to large-scale interaction between the ionized gas and of stellar winds. A process that may dominate the kinematics of these complex and huge nebulae. This suggestion is based on the results obtained by Mesa-Delgado et al. (2014) and Esteban et al. (2016), who determined the ADF(O$^{2+}$) in several Galactic ring nebulae associated with evolved massive stars. These objects show mean ADF(O$^{2+}$) between 0.38 and 0.50 dex, values larger than the ADFs of normal {\hii} regions with the same O abundance but similar to the high ADF(O$^{2+}$) values found in star-forming dwarf galaxies.

\acknowledgments CE thanks Oli Dors Jr. and the rest of the members of the LOC for their kind invitation and financial support to participate in this enjoyable workshop. LTSC is supported by the FPI Program by MINECO under grant AYA2011-22614. This work was also partially funded by MINECO under grant AYA2015-65205-P.

\end{document}